\documentclass[msmath,amssymb,aps,pra,twocolumn,epsfig,showpacs,bibliography,
lengthcheck,superscriptaddress]{revtex4-1}

\usepackage{graphicx}
\usepackage{dcolumn}
\usepackage{bm}
\usepackage{hyperref}
\usepackage{epstopdf}
\usepackage{subcaption}
\usepackage{caption}
\usepackage{amsmath}
\begin{document}
\title{Electric quadrupole shifts of the precession frequencies of $^{131}$Xe atoms in rectangular cells}
\author{Y.-K. Feng}
\author{S.-B. Zhang}
\author{Z.-T. Lu}
\affiliation{Hefei National Laboratory for Physical Sciences at the Microscale, CAS Center for Excellence in Quantum Information and Quantum Physics, University of Science and Technology of China, Hefei 230026, China}
\affiliation{Department of Modern Physics, University of Science and Technology of China, Hefei 230026, China}
\author{D. Sheng}
\email{dsheng@ustc.edu.cn}
\affiliation{Hefei National Laboratory for Physical Sciences at the Microscale, CAS Center for Excellence in Quantum Information and Quantum Physics, University of Science and Technology of China, Hefei 230026, China}
\affiliation{Department of Precision Machinery and Precision Instrumentation, Key Laboratory of Precision Scientific Instrumentation of Anhui Higher Education Institutes, University of Science and Technology of China, Hefei 230027, China}

\begin{abstract}
We study an atomic comagnetometer design based on the spin precessions of $^{129}$Xe and $^{131}$Xe atoms in glass cells. The quadrupole splittings in the precession spectrum of $^{131}$Xe are fully resolved, allowing a precise determination of the magnetic-dipole precession frequency. The transverse asymmetry of quadrupole interactions, due to both the geometry and surface properties of the cell, characterized by a non-zero asymmetry parameter $\eta$, modifies the dependence of the quadrupole splittings on the relative orientation between the cell axes and the bias magnetic field, and lead to additional corrections in the precession frequencies of $^{131}$Xe atoms. We examine these effects both theoretically and experimentally, and develop methods to quantify and control such shifts.
\end{abstract}

\maketitle

\section{Introduction}
A noble-gas comagnetometer consists of two kinds of atoms, whose nuclear spin precession frequencies are inter-compared to cancel any drifts and noises due to changes in the bias magnetic field. In a hybrid system mixed with alkali atoms, noble-gas atoms are polarized~\cite{bouchiat1960,walker1997} and probed~\cite{grover1978} through their interactions with alkali atoms. Such systems have been widely used in precision measurements of both fundamental physics~\cite{bear2000,rosenberry2001,tullney2013,bulatowicz2013,kover2015,limes18} and inertial sensing~\cite{donley2013,walker16}.

The coupling between the nuclear magnetic dipole moment $\bm{\mu}$ and the magnetic field $\mathbf{B}$, under the Hamiltonian
\begin{equation}
\label{eq:Hd}
H_D=-\bm{\mu}\cdot\mathbf{B},
\end{equation}
determines the magnetic-dipole precession frequency $\omega=\gamma{B}$, where $\gamma$ is the nuclear gyromagnetic ratio. In an ideal comagnetometer, two isotopes of both nuclear spin 1/2, denoted as $a$ and $b$, occupy the same spatial region. Consequently, the ratio of their precession frequencies,
\begin{equation}
\label{eq:comag}
{\omega_a}/{\omega_b}={\gamma_a}/{\gamma_b},
\end{equation}
is independent of the external field. There are indeed two and only two stable noble-gas isotopes with nuclear spin 1/2: $^3$He and $^{129}$Xe. A $^3$He-$^{129}$Xe comagnetometer system mixed with alkali atoms suffers from a systematic effect due to the large difference between  helium and xenon in their atomic sizes and collision properties with the alkali atoms. Consequently, $^3$He and $^{129}$Xe atoms experience different effective magnetic fields generated by the polarized alkali atoms~\cite{schaefer1989,romalis1998,ma2011}. Various schemes have been developed to solve this problem~\cite{sheng14a,kover2015,limes18}, but the strong modulation fields introduced in these methods present potential problems on the stability of the system.

Besides $^3$He, the stable isotope $^{131}$Xe ($I$=3/2) can also be combined with $^{129}$Xe to form a comagnetometer. Here, the two xenon isotopes have nearly identical collisional properties with alkali atoms~\cite{bulatowicz2013}, thus suppressing the aforementioned systematic effect. However, with a nuclear spin of $3/2$, $^{131}$Xe has not only a nuclear magnetic dipole moment, but also a nuclear electric quadrupole moment that couples to the external electric field gradient, during the dwelling time ($\sim10^{-11}$~s) of the atoms on the cell surfaces~\cite{kwon81,wu1987,wu1990}.

\begin{figure}[htb]
\includegraphics[width=2.0in]{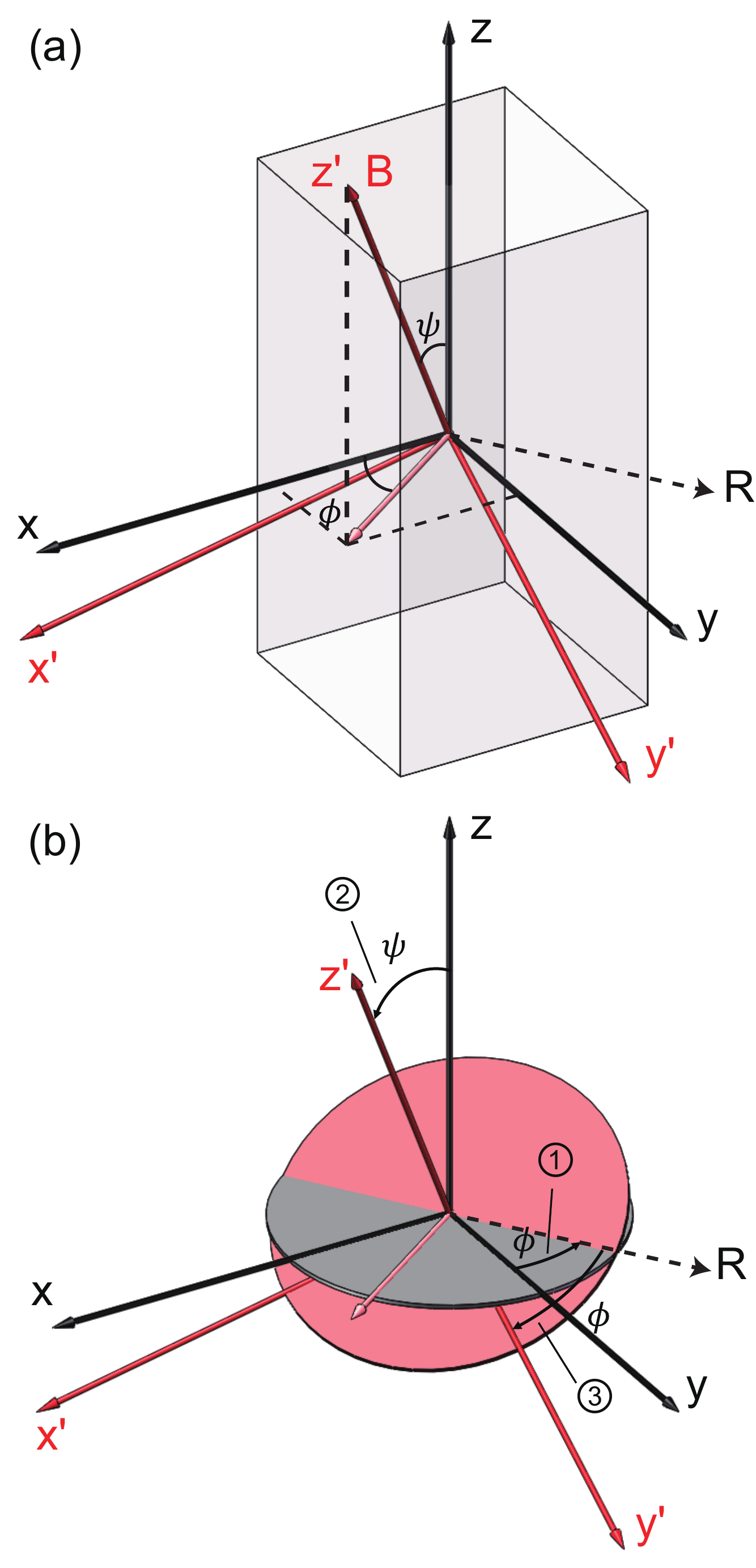}
\caption{\label{fig:Bangle}(Color online) (a) A rectangular cell, whose principal axes form the $xyz$ coordinate, in an arbitrarily oriented magnetic field $\bm{B}$. The $x'y'z'$ coordinate, defined by the quantization axis, can be obtained by rotating the $xyz$ coordinate around the $R$ axis that is perpendicular to the $zz'$ plane.  (b) The transformation from the $xyz$ coordinate to the $x'y'z'$ coordinate can also be achieved by the Euler rotations, where the circled numbers in the plot denote the sequence of the rotations.}
\end{figure}

Figure~\ref{fig:Bangle}(a) shows two coordinate systems that are set up in this paper: the $xyz$ system defined by the principal axes of the rectangular cell, and the $x'y'z'$ system, where the $z'$ axis is the quantization axis. When the two systems coincide with each other, the Hamiltonian for the electric quadrupole interaction is~\cite{slichter90}
\begin{equation}
\label{eq:Hq}
H_{Q0}=\frac{eQ}{4I(2I-1)}[V_{zz}(3\bm{I}_z^2-\bm{I}^2)+(V_{xx}-V_{yy})(\bm{I}_x^2-\bm{I}_y^2)],
\end{equation}
where $Q$ is the electric quadrupole moment of the atom, and $V_{ii}$ is the second-order derivative of the electric potential along the $i$ direction. Because the interaction happens only when the atoms dwell on the cell surfaces, experimentally observed quadrupole effects in $^{131}$Xe precession spectrum are time-averaged, and can be described by an effective Hamiltonian which is equal to $H_{Q0}$ in Eq.~(\ref{eq:Hq}) times a small coefficient $p$~\cite{kwon81}. By defining $q=V_{zz}/e$ and the asymmetry parameter $\eta=(V_{xx}-V_{yy})/{V_{zz}}$, Eq.~(\ref{eq:Hq}) becomes:
\begin{equation}
\label{eq:Hq2}
H_{Q0}=\frac{pe^2qQ}{4I(2I-1)}[3\bm{I}_z^2-\bm{I}^2+\eta(\bm{I}_x^2-\bm{I}_y^2)].
\end{equation}

The combination of the magnetic dipole and electric quadrupole interactions leads to a triplet structure in the precession spectrum of $^{131}$Xe. To recover the ideal comagnetometer working condition in Eq.~\eqref{eq:comag} for $^{131}$Xe atoms, it is important to understand and control the quadrupole splittings among the triplet in the precession spectrum. For an arbitrary magnetic field orientation, $H_{Q0}$ is transformed into $H_Q$ using the rotational properties of the angular momentum~\cite{sakurai2013}. In the case of strict $x-y$ symmetry, $\eta=0$, $H_Q$ shows a 90$^\circ$-rotation symmetry around the $z$ axis~\cite{venema1992,donley2009}. In practice, however, the 90$^\circ$-rotation symmetry of the quadrupole interaction is often degraded by imperfect cell conditions, for example, the addition of a cell stem and non-uniform deposition of alkali atoms on the cell surfaces. This asymmetry leads to additional frequency shifts, and complicates the combination method used to extract the magnetic-dipole precession frequency of $^{131}$Xe~\cite{bulatowicz2013}.




In this work, we examine the comagnetometer system of $^{129}$Xe-$^{131}$Xe mixed with Rb atoms in rectangular cells with a square cross section, and focus on understanding the quadrupole shifts and systematics of the precession spectrum of $^{131}$Xe when the asymmetry parameter $\eta$ in the quadrupole interaction can not be neglected. Following this introduction, Sec.~II presents the theoretical formulation of the quadrupole splittings, Sec.~III describes the experiment apparatus and methods, Sec.~IV shows the measurement results, and Sec.~V concludes the paper.

\section{Modeling the quadrupole splittings}

When the $xyz$ coordinate coincides with $x'y'z'$ coordinate, the quadrupole interaction Hamiltonian for $^{131}$Xe atoms, previously expressed in Eq.~\eqref{eq:Hq2}, can be written in the matrix form
\begin{eqnarray}~\label{eq:Hq0}
H_{Q0}=\frac{pe^2qQ}{12}\left(
\begin{array}{cccc}
 3 & 0 & \sqrt{3}\eta & 0 \\
 0 & -3 & 0 & \sqrt{3}\eta \\
 \sqrt{3}\eta & 0 & -3 & 0 \\
 0 & \sqrt{3}\eta & 0 & 3 \\
\end{array}
\right).
\end{eqnarray}

For an arbitrary $\mathbf{B}$, characterized by its zenith angle $\psi$ and azimuth angle $\phi$ shown in Fig.~\ref{fig:Bangle}(a), the $xyz$ and $x'y'z'$ coordinates are connected by a single rotation around the $R$ axis, which is perpendicular to the $zz'$ plane. Such a rotation can also be described by three successive Euler rotations shown in  Fig.~\ref{fig:Bangle}(b), with the Euler angles of $(\phi,\psi,-\phi)$. Then the interaction Hamiltonian in Eq.~\eqref{eq:Hq0} is transformed as
\begin{equation}~\label{eq:Hq3}
H_Q=\mathcal{D}_{\frac{3}{2}}(\phi,\psi,-\phi) H_{Q0}\mathcal{D}^{-1}_{\frac{3}{2}}(\phi,\psi,-\phi),
\end{equation}
where $\mathcal{D}_{j}$ is the Wigner D matrix for the angular momentum $\bm{j}$.

The complete Hamitonian for $^{131}$Xe, including both the magnetic dipole and electric quadrupole interactions, is
\begin{widetext}
\begin{eqnarray}
&&H=\hbar\omega\left(\begin{array}{cccc}
\frac{3}{2} &0 & 0 & 0 \\
 0 & \frac{1}{2} & 0 & 0 \\
 0 & 0 & -\frac{1}{2} & 0 \\
 0 &0 &0 & -\frac{3}{2} \\
\end{array}\right)+\frac{pe^2qQ}{12}\left(
\begin{array}{cccc}
 a & b & c & 0 \\
 b^* & -a & 0 & c \\
 c^* & 0 & -a & -b \\
 0 & c^* & -b^* & a \\
\end{array}
\right)~\label{eq:H},\\
&&a=\frac{3}{2}\left[3 \cos ^2(\psi )-1+\eta  \sin ^2(\psi ) \cos (2 \phi )\right],\\
&&b=\sqrt{3} \sin (\psi )e^{-i\phi}\left\{3 \cos (\psi )-\eta  [\cos (\psi ) \cos (2 \phi )+i \sin (2 \phi )]\right\},\\
&&c=\frac{ \sqrt{3}}{2}e^{-2i\phi} \left\{3 \sin ^2(\psi )+\eta \left[\left(\cos ^2(\psi )+1\right) \cos (2 \phi )+2 i \cos (\psi ) \sin (2 \phi )\right]\right\}.
\end{eqnarray}
\end{widetext}

To simplify the expression in the above equations, we introduce a  parameter $\omega_q=pe^2qQ/2\hbar$. For experimental conditions of $^{131}$Xe atoms studied in this paper, $\omega_q/2\pi\sim$50~mHz, and the magnetic-dipole precession frequency $\omega/2\pi\sim$3~Hz. Therefore, we can treat the quadrupole interaction as a perturbation to the dipole interaction. In the second order approximation of quadrupole interactions, the eigenvalues of $H$ in Eq.~(\ref{eq:H}) are
\begin{eqnarray}
&&E_1=\frac{3}{2}\hbar\omega+\frac{\hbar\omega_q}{6}a+\frac{\hbar\omega_q^2}{36\omega}\left(|b|^2+\frac{|c|^2}{2}\right),\\
&&E_2=\frac{1}{2}\hbar\omega-\frac{\hbar\omega_q}{6}a-\frac{\hbar\omega_q^2}{36\omega}\left(|b|^2-\frac{|c|^2}{2}\right),\\
&&E_3=-\frac{1}{2}\hbar\omega-\frac{\hbar\omega_q}{6}a+\frac{\hbar\omega_q^2}{36\omega}\left(|b|^2-\frac{|c|^2}{2}\right),\\
&&E_4=-\frac{3}{2}\hbar\omega+\frac{\hbar\omega_q}{6}a-\frac{\hbar\omega_q^2}{36\omega}\left(|b|^2+\frac{|c|^2}{2}\right).
\end{eqnarray}

This leads to the three observed transition lines in the precession spectrum of $^{131}$Xe atoms:
\begin{eqnarray}
&&\label{eq:w+}\hbar\omega_+=E_1-E_2=\hbar\omega+\frac{\hbar\omega_q}{3}a+\frac{\hbar\omega_q^2}{18\omega}|b|^2,\\
&&\label{eq:w0}\hbar\omega_0=E_2-E_3=\hbar\omega-\frac{\hbar\omega_q^2}{18\omega}\left(|b|^2-\frac{|c|^2}{2}\right),\\
&&\label{eq:w-}\hbar\omega_-=E_3-E_4=\hbar\omega-\frac{\hbar\omega_q}{3}a+\frac{\hbar\omega_q^2}{18\omega}|b|^2.
\end{eqnarray}
We define the difference between $\omega_+$ and $\omega_-$ in the equations above as the quadrupole splitting $\Omega$:
\begin{eqnarray}~\label{eq:omega}
\Omega&=&\omega_+-\omega_-=\frac{2}{3}\omega_qa\nonumber\\
&=&\omega_q[3\cos^2(\psi)-1+\eta\sin^2(\psi)\cos(2\phi)].
\end{eqnarray}
Note that, when $\eta=0$, $\Omega$ is only dependent on the zenith angle $\psi$ between $\mathbf{B}$ and the $z$ axis in the second order approximation of the quadrupole interactions, same as the cases of cylindrical and spherical cells~\cite{kwon81,wu1987}. Another indication from Eq.~\eqref{eq:omega} is that the dependence of $\Omega$ on $\eta$ can be removed by combining the results from two different magnetic field orientations or choosing $\phi=\pi/4$:
\begin{eqnarray}~\label{eq:w45}
\frac{\Omega(\psi,\phi)+\Omega(\psi,\frac{\pi}{2}+\phi)}{2}&=&\Omega(\psi,\pi/4)\nonumber\\
&=&\omega_q\left[3\cos^2(\psi)-1\right].
\end{eqnarray}

\section{APPARATUS AND METHODS}
\begin{figure}[hbt]
\includegraphics[width=3in]{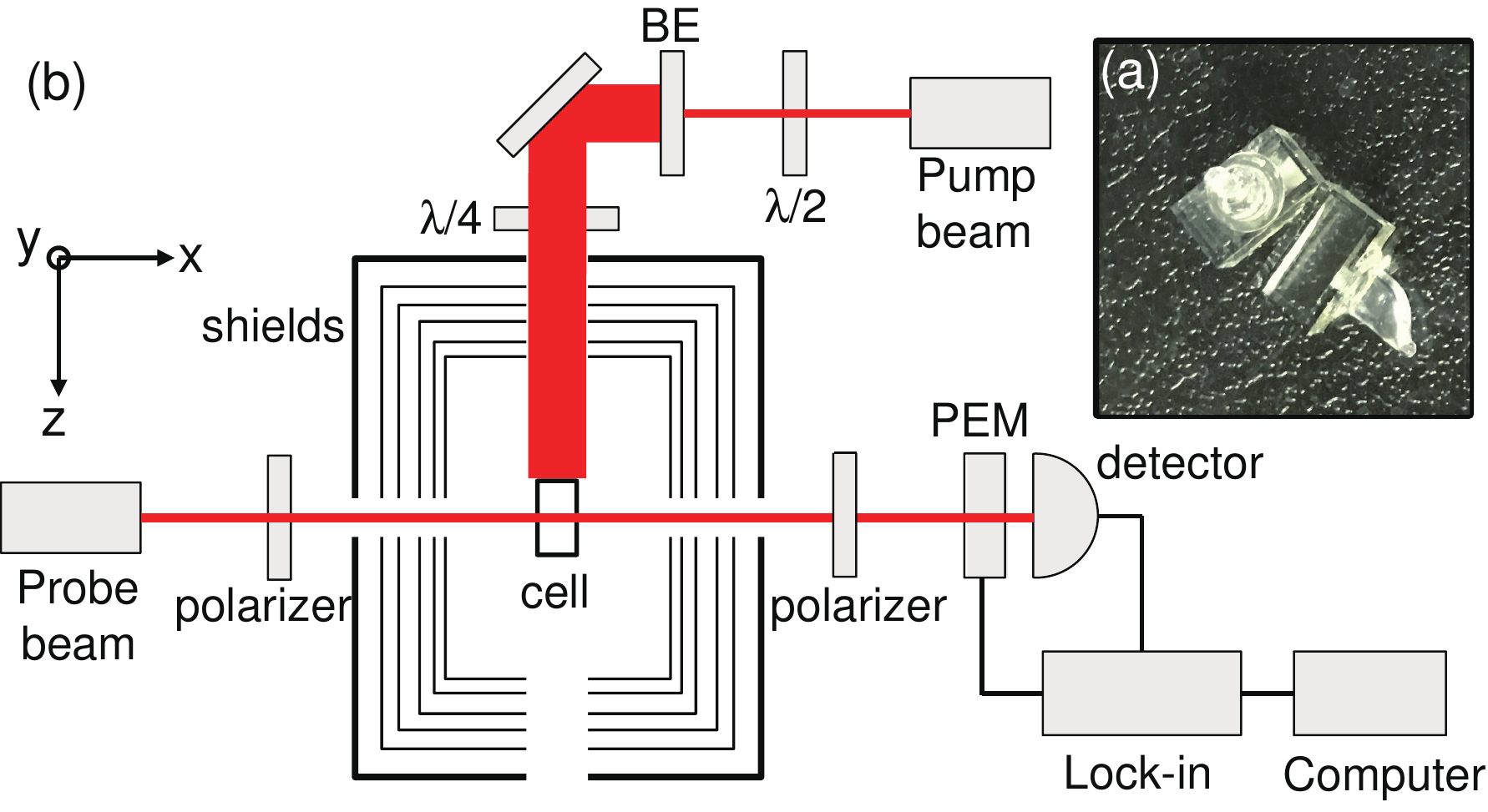}
\caption{\label{fig:setup}(Color online) (a) Atomic cells used in the experiment. (b) Experiment setup (field coils not shown in the figure), BE: beam expander.}
\end{figure}

In the experiment, we use rectangular cells made of Pyrex glass  (Fig.~\ref{fig:setup}(a)). Each cell has an inner dimension of 9~mm$\times$4~mm$\times$4~mm. The cell is loaded with Rb atoms of natural abundances and the following gases: 15 Torr of $^{131}$Xe, 4 Torr of $^{129}$Xe, 5 Torr of H$_2$, and 400 Torr of N$_{2}$. The cell sits inside an oven, heated by ac currents with a frequency of 70~kHz through twisted resistance wires, and placed inside a five-layer cylindrical mu-metal magnetic shields. We align the long axis of the cell with the longitudinal axis of the shields ($z$ axis in Fig.~\ref{fig:setup}(b)). The magnetic fields inside the shields are controlled by three sets of coils, with a set of solenoid coils for the longitudinal field, and two sets of cosine-theta coils for the transverse field~\cite{venemathesis}.

A circularly polarized laser beam passes through the cell along the $z$ axis. With a beam diameter of 1~cm and a beam intensity of 50~mW/cm$^2$, this pump beam is on resonance with the D1 transitions of both Rb isotopes, whose full linewidths are pressure broadened to 9~GHz~\cite{romalis1997}. A linearly polarized probe beam 5~GHz red detuned from the Rb D1 line, with a beam diameter of 2~mm and a beam power of 3~mW, passes through the cell along the $x$ axis. The pump and probe beams are generated from two separate distributed-Bragg-reflector laser diodes.  We detect the Rb magnetometer signals by analyzing the probe beam polarization with a polarimeter, which consists of a photo-elastic modulator (PEM) modulating the light polarization at a frequency of 50~kHz, a set of cross-polarizers (one before the cell and one after the cell in Fig.~\ref{fig:setup}), and a photodiode detector. The detector signal is demodulated by a lock-in amplifier.

Each experiment cycle lasts 180 s, consisting of three stages. In the first stage, the polarized Rb atoms exchange polarization with the Xe atoms over 60 seconds under a bias field $\mathbf{B}_0=$9~mG in the $z$ direction. At the beginning of the second stage, a short $\pi/2$ pulse is applied to simultaneously tip the polarizations of both $^{129}$Xe and $^{131}$Xe to the $xy$ plane. Following this pulse, the bias magnetic field is adjusted to $\mathbf{B}_1$, with $|\mathbf{B}_1|=|\mathbf{B}_0|$, whose direction is varied as to study the effect of the magnetic field orientation on Xe precession spectrum. Xe atoms precess around $\mathbf{B}_1$ for 90 seconds while their spins are monitored by the probe beam through the Rb magnetometer. In the third stage, preparing for the next experiment cycle, the bias field is changed back to the $\mathbf{B}_0$, and a 30 s long field gradient pulse ($dB_z/dz$) is applied to depolarize completely Xe atoms. The pump and probe beams are kept on throughout the cycle.

\begin{figure}[hbt]
\includegraphics[width=3in]{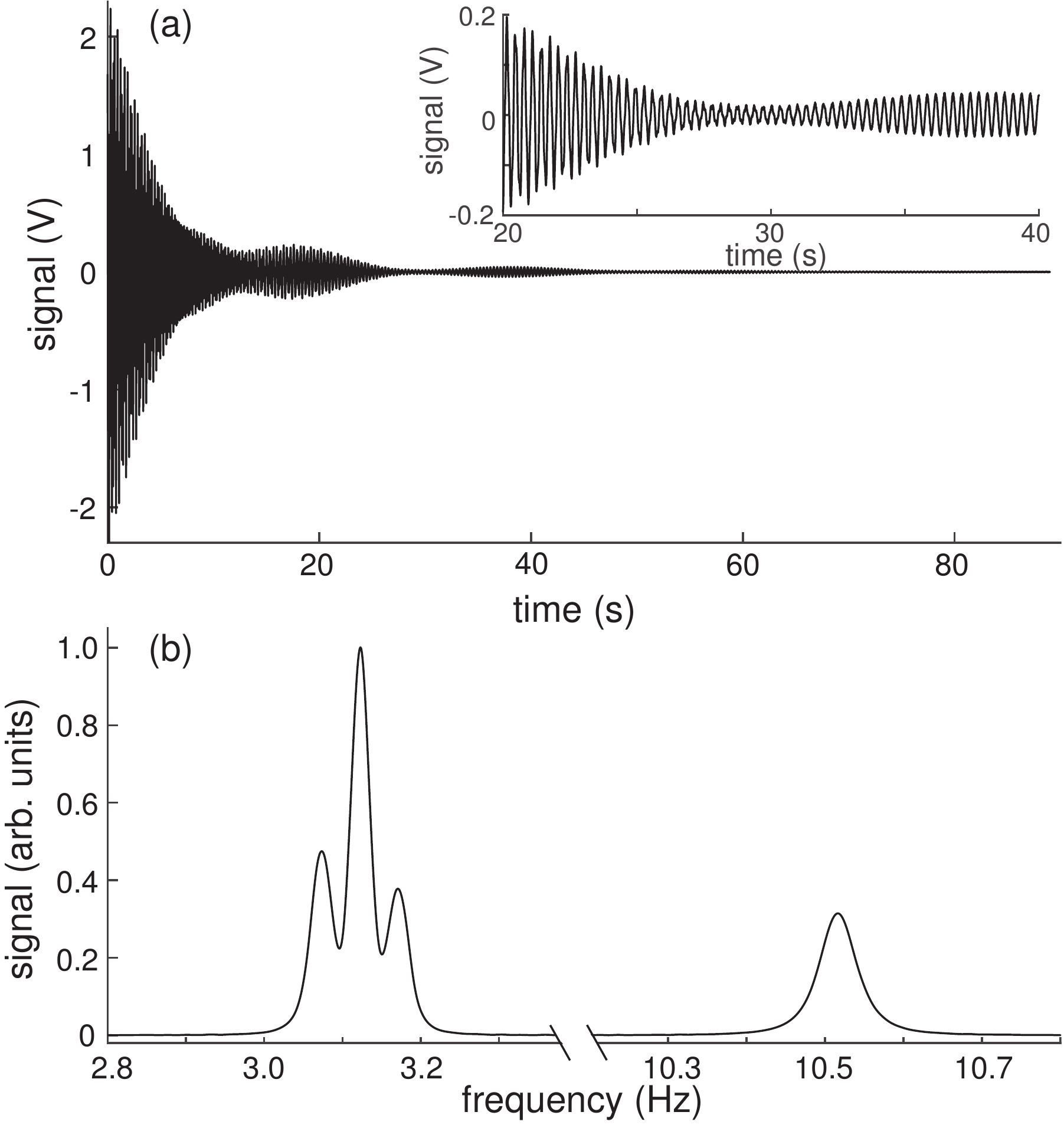}
\caption{\label{fig:sig1}(a) A typical precession signal of Xe atoms probed by the Rb magnetometer, with an expanded view of the data in the inset. (b) The amplitude spectrum of the signal in plot (a) using the Hanning window.}
\end{figure}

Figure~\ref{fig:sig1}(a) shows a typical plot of the Xe precession signal recorded during the second stage of an experimental cycle at the cell temperature of 383~K. The Fourier transform of the experimental data shows four peaks in Fig.~\ref{fig:sig1}(b). While the Larmor frequency of $^{129}$Xe is at 10.5 Hz, the other three peaks near 3.1 Hz form the quadrupole-split spectrum of $^{131}$Xe.
We use four exponential-decay-oscillation functions to fit the data:
\begin{equation}~\label{eq:fit}
y=\sum_{i=1}^4a_ie^{-(t-t_0)/\tau_i}\sin[2\pi f_i(t-t_0+\varphi_i)]+b,
\end{equation}
The typical fitting error is 1 $\mu$Hz for the $^{129}$Xe precession frequency, and are 4~$\mu$Hz and 15~$\mu$Hz for the central peak and sidebands of $^{131}$Xe atoms, respectively. The quadrupole splittings of $^{131}$Xe atoms are extracted from the fitting results.

\section{Results and Discussion}
\begin{figure}
\includegraphics[width=3in]{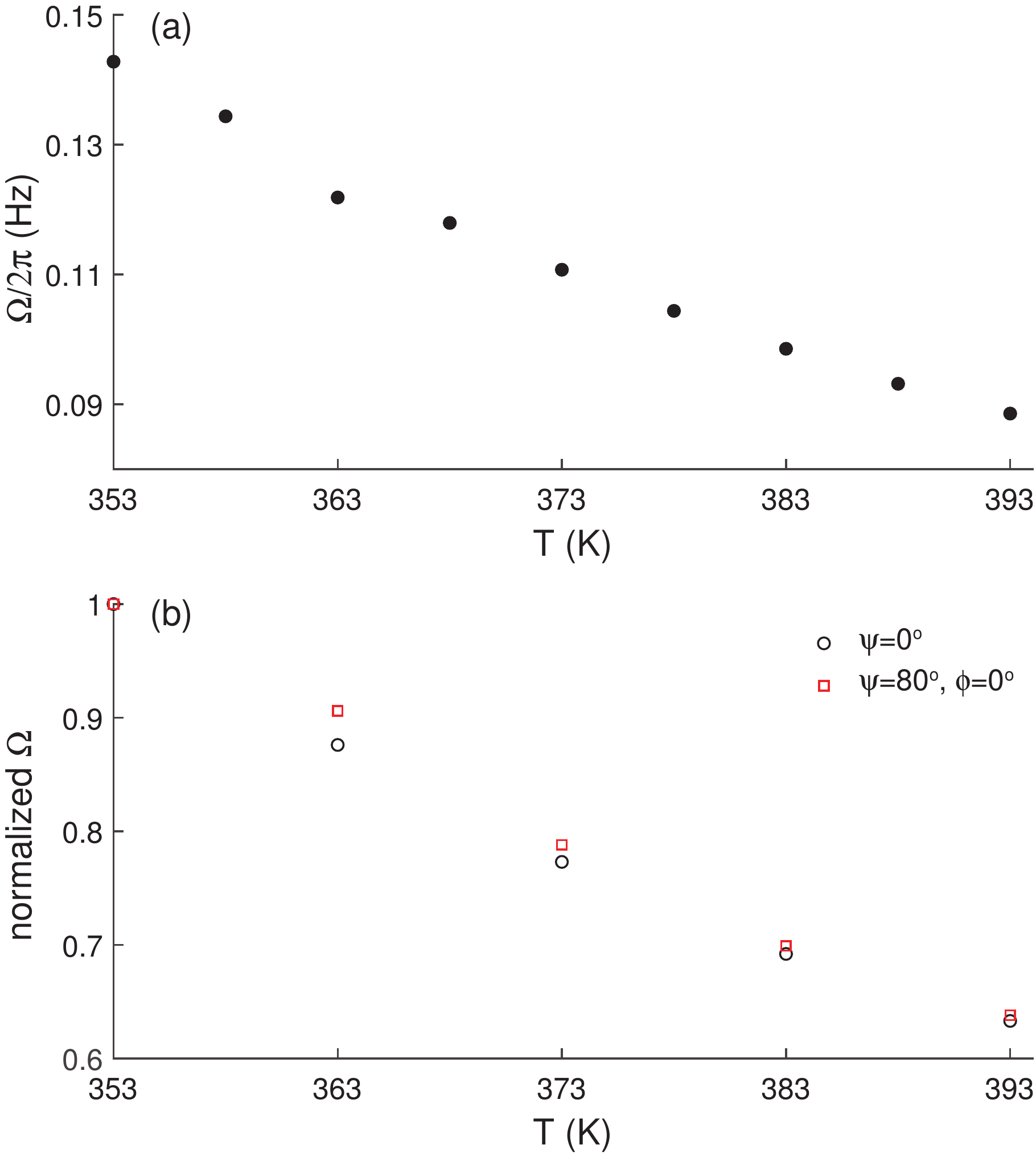}
\caption{(Color online)\label{fig:QST}(a) Temperature dependence of the quadrupole splitting $\Omega$ on $T$, with the magnetic field $\mathbf{B}$ along the $z$ axis. (b) $\Omega(T)/\Omega(T_0=353~K)$ for two different $\mathbf{B}$ orientations.}
\end{figure}

We first characterize the properties of the quadrupole interactions between the $^{131}$Xe atoms and the cell surfaces. As discussed in Ref.~\cite{wu1987}, the dependence of the quadrupole splittings on the cell wall temperature is described by the relation:
\begin{equation}~\label{eq:qsT}
\Omega\propto e^{-E_a/k_BT},
\end{equation}
where $E_a$ is the adsorption potential of the cell walls acting on the $^{131}$Xe atoms. Figure~\ref{fig:QST}(a) shows the measured temperature dependence of $\Omega$ when the magnetic field is along the $z$ axis. Using Eq.~\eqref{eq:qsT}, we extract the adsorption energy $E_a=-0.14\pm0.01$~eV. This result is consistent with previously measured activation energy of cured Pyrex cells filled with H$_2$ gas~\cite{wu1990}. Figure~\ref{fig:QST}(b) shows that the normalized results of $\Omega(T)/\Omega(T_0=353~K)$ for two different magnetic field directions agree with each other, confirming the expectation that the relation between $\Omega$ and the direction of the bias field is independent of the cell temperature. We then fix the cell temperature at $T=383$~K for the rest of the study.

To probe the dependence of the $^{131}$Xe quadrupole splittings on the magnetic field, we performed measurements on two atomic cells, which have the same conditions except the distribution of Rb atoms inside the cells. Cell \#1 has Rb droplets accumulated on one of its inner surfaces, while most the Rb atoms of cell \#2 are chased into the cell tip. For each cell, we study three cases: $\mathbf{B}_1$ in the second stage of the experiment stays in the $xz$ plane ($\phi=0^\circ$), $yz$ plane ($\phi=90^\circ$), and the plane of $\phi=45^\circ$. In each case, we scan the zenith angle $\psi$ of $\mathbf{B}_1$ using the field coils.

\begin{figure}[hbt]
\includegraphics[width=3in]{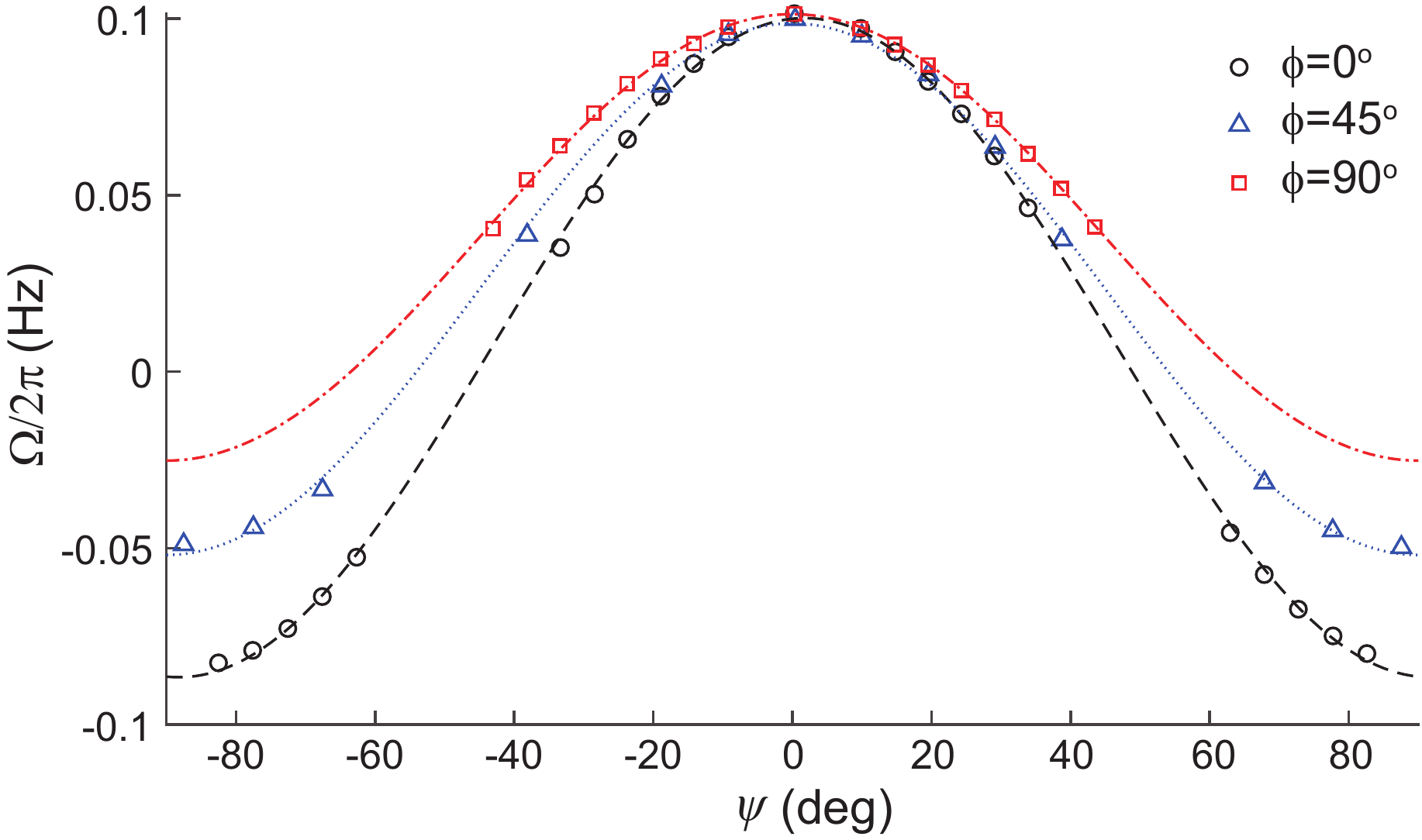}
\caption{\label{fig:quadang} (Color Online) The data points  are the experiment results of quadrupole splittings of $^{131}$Xe atoms in cell \#1 as a function of the magnetic field orientation angle $\psi$, for the cases of $\phi=0^\circ$, $45^\circ$ and $90^\circ$. The lines are the corresponding fitting results using Eq.~\eqref{eq:omega}.}
\end{figure}

Figure~\ref{fig:quadang} shows the experimental data of cell \#1, together with the fitting lines using Eq.~\eqref{eq:omega}, where we leave $\omega_q$, $\eta$, and the offset of $\psi$ ($\psi_0$) as free parameters. The data for the case of $\phi=45^\circ$ gives a fitting result of $\eta\cos\phi=-0.05\pm0.02$, consistent with the prediction of Eq.~\eqref{eq:w45} that the quadrupole splitting is insensitive to $\eta$ in this case.  Data sets of $\phi=0^\circ$ and $90^\circ$ show clear evidences of asymmetrical quadrupole interactions. Using the combined fitting results of both data sets, we extracted $\omega_q/2\pi=50.4\pm0.3$~mHz and $|\eta|=0.62\pm0.1$ for cell \#1. There are two possible origins for this surprisingly large asymmetry parameter in cell \#1. One is the geometric asymmetry due to the cell stem, and the other is that there are several Rb droplets accumulated on one of inner surfaces~\cite{volk1979}.  To quantify each effect, we performed similar measurements on cell \#2, and obtained $\omega_q/2\pi=62.4\pm0.1$~mHz and $|\eta|=0.22\pm0.02$. This suggests that Rb deposition on the cell surfaces can be a significant factor, possibly even the dominant one for the asymmetry parameter.


As discussed in the introduction, it is often required in precision measurements to extrapolate the magnetic-dipole precession frequency ($\omega$) of $^{131}$Xe from the measured $\omega_+$, $\omega_0$, and $\omega_-$ in Eqs.~(\ref{eq:w+}-\ref{eq:w-}). Compared with $\omega_\pm$, $\omega_0$ is closer to $\omega$ because it contains only a quadratic term of $\omega_q$. Two other combinations have been used in the literature:
\begin{widetext}
\begin{eqnarray}
&&\omega_1=\frac{\omega_++\omega_-}{2}=\omega+\frac{\omega_q^2}{6\omega}\sin^2(\psi)\left\{\left[3-\eta\cos(2\phi)\right]^2\cos^2(\psi)+\eta^2\sin^2(2\phi)\right\}\label{eq:w1}\\
&&\omega_2=\frac{\omega_++\omega_-+2\omega_0}{4}=\omega+\frac{\omega_q^2}{96\omega}\left\{\left[3\sin^2(\psi)+\eta(\cos^2(\psi)+1)\cos(2\phi)\right]^2+4\eta^2\cos^2(\psi)\sin^2(2\phi)\right\}\label{eq:w2}
\end{eqnarray}
\end{widetext}

It is good practice in experiments to align the magnetic field direction as close as possible along the $z$ axis so that $\psi$ is near zero. Consequently, $\omega_{0,1,2}$ are simplified to:
\begin{eqnarray}
&&\begin{split}
\omega_0=\omega-\frac{\omega_q^2}{\omega}\left\{\left[\frac{3}{2}-\frac{5}{4}\eta\cos(2\phi)+\frac{\eta^2}{6}\right]\psi^2\right.\\
\left.-\frac{\eta^2}{12}-\frac{3}{16}\psi^4\right\},
\end{split}\label{eq:shift0}\\
\label{eq:shift1}&&\omega_1=\omega+\frac{\omega_q^2}{6\omega}\psi^2[9-6\eta\cos(2\phi)+\eta^2],\\
\label{eq:shift2}&&\omega_2=\omega+\frac{\omega_q^2}{96\omega}[4\eta^2+12\eta\psi^2\cos(2\phi)+9\psi^4].
\end{eqnarray}

In past works, the x-y asymmetry properties were often neglected~\cite{kwon81,wu1987,Majumder1990,venema1992,donley2009}, as if the parameter $\eta$ were assumed to be zero. Under such an assumption, $\omega_2$ is preferred because it deviates from $\omega$ only by a small term of $\psi^4$. However, as demonstrated in this paper, $\eta$ should not be considered zero, as its size can be significantly larger than $\sin\psi$. In the two cells examined in this work, $\psi=0.01$, much less than the $\eta$ of each cell. Taking cell \#1 for example ($|\eta|=0.6$, $\omega_q/2\pi=50$~mHz, $\psi=0.01$, and $\omega/2\pi=3$~Hz), the asymmetrical quadrupole interaction adds a frequency shift of 12~$\mu$Hz to $\omega_2$ and 24~$\mu$Hz to $\omega_0$. A $1\%$ change of $\eta$ during an experiment run, possibly caused by a redistribution of Rb deposition inside the cell, would lead to a shift of 250~nHz to $\omega_2$ and 500~nHz to $\omega_0$. $\omega_1$ can be a much better choice because its asymmetrical quadrupole term is proportional to $\psi^2$. In the example of cell \#1, the asymmetrical quadrupole shift is only 100~nHz, two orders of magnitude smaller than those of $\omega_0$ and $\omega_2$.

\section{Conclusion}

In summary, we have studied the electric quadrupole splittings of $^{131}$Xe atoms in rectangular cells with non-zero asymmetry parameters, which introduce additional quadrupole shifts to the $^{131}$Xe precession frequencies. These effects are caused not only by the asymmetries in the geometric properties, but also in the surface properties of the cells. In particular, we find that non-uniform Rb covering of the cell can induce a significantly nonzero $\eta$ parameter. These effects are expected to be present even if the cells are in cylindrical, spherical or cubic shapes.

In certain types of experiments where additional modulation parameters are available, for example the spin-gravity experiment~\cite{venema1992}, the quadrupole shifts introduced by the asymmetry parameter can be cancelled by comparing successive runs of flipped parameters. However, it is an important systematic effect when the accuracy of the $^{131}$Xe magnetic-dipole precession frequency is required, such as extracting the absolute value of $\gamma_\mathrm{Xe129}/\gamma_\mathrm{Xe131}$.

In addition to measuring $\eta$ and correcting the asymmetric quadrupole shifts with a high precision as demonstrated in this paper, one can suppress the asymmetric quadrupole shifts as follows. First, one can increase the bias field strength, thus reducing $\omega_q/\omega$. Second, one can reduce $\eta$, for example, by reducing the stem size (or even removing the stem~\cite{limes2019}) and by chasing excess Rb into the stem. Third, one can use $\omega_1$ in Eq.~\eqref{eq:shift1} to approximate $\omega$~\cite{bulatowicz2013}, which takes advantage of the small factor $\psi^2$ in the expression.

\section*{Acknowledgements}
This work was supported by Natural Science Foundation of China (Grant No.11774329 and 11974329) and Key Research Program of Frontier Sciences, CAS (NO. XDB21010200).


\begin{thebibliography}{27}%
\makeatletter
\providecommand \@ifxundefined [1]{%
 \@ifx{#1\undefined}
}%
\providecommand \@ifnum [1]{%
 \ifnum #1\expandafter \@firstoftwo
 \else \expandafter \@secondoftwo
 \fi
}%
\providecommand \@ifx [1]{%
 \ifx #1\expandafter \@firstoftwo
 \else \expandafter \@secondoftwo
 \fi
}%
\providecommand \natexlab [1]{#1}%
\providecommand \enquote  [1]{``#1''}%
\providecommand \bibnamefont  [1]{#1}%
\providecommand \bibfnamefont [1]{#1}%
\providecommand \citenamefont [1]{#1}%
\providecommand \href@noop [0]{\@secondoftwo}%
\providecommand \href [0]{\begingroup \@sanitize@url \@href}%
\providecommand \@href[1]{\@@startlink{#1}\@@href}%
\providecommand \@@href[1]{\endgroup#1\@@endlink}%
\providecommand \@sanitize@url [0]{\catcode `\\12\catcode `\$12\catcode
  `\&12\catcode `\#12\catcode `\^12\catcode `\_12\catcode `\%12\relax}%
\providecommand \@@startlink[1]{}%
\providecommand \@@endlink[0]{}%
\providecommand \url  [0]{\begingroup\@sanitize@url \@url }%
\providecommand \@url [1]{\endgroup\@href {#1}{\urlprefix }}%
\providecommand \urlprefix  [0]{URL }%
\providecommand \Eprint [0]{\href }%
\providecommand \doibase [0]{http://dx.doi.org/}%
\providecommand \selectlanguage [0]{\@gobble}%
\providecommand \bibinfo  [0]{\@secondoftwo}%
\providecommand \bibfield  [0]{\@secondoftwo}%
\providecommand \translation [1]{[#1]}%
\providecommand \BibitemOpen [0]{}%
\providecommand \bibitemStop [0]{}%
\providecommand \bibitemNoStop [0]{.\EOS\space}%
\providecommand \EOS [0]{\spacefactor3000\relax}%
\providecommand \BibitemShut  [1]{\csname bibitem#1\endcsname}%
\let\auto@bib@innerbib\@empty
\bibitem [{\citenamefont {Bouchiat}\ \emph {et~al.}(1960)\citenamefont
  {Bouchiat}, \citenamefont {Carver},\ and\ \citenamefont
  {Varnum}}]{bouchiat1960}%
  \BibitemOpen
  \bibfield  {author} {\bibinfo {author} {\bibfnamefont {M.~A.}\ \bibnamefont
  {Bouchiat}}, \bibinfo {author} {\bibfnamefont {T.~R.}\ \bibnamefont
  {Carver}}, \ and\ \bibinfo {author} {\bibfnamefont {C.~M.}\ \bibnamefont
  {Varnum}},\ }\href {\doibase DOI 10.1103/PhysRevLett.5.373} {\bibfield
  {journal} {\bibinfo  {journal} {Phys. Rev. Lett.}\ }\textbf {\bibinfo
  {volume} {5}},\ \bibinfo {pages} {373} (\bibinfo {year} {1960})}\BibitemShut
  {NoStop}%
\bibitem [{\citenamefont {Walker}\ and\ \citenamefont
  {Happer}(1997)}]{walker1997}%
  \BibitemOpen
  \bibfield  {author} {\bibinfo {author} {\bibfnamefont {T.~G.}\ \bibnamefont
  {Walker}}\ and\ \bibinfo {author} {\bibfnamefont {W.}~\bibnamefont
  {Happer}},\ }\href {\doibase DOI 10.1103/RevModPhys.69.629} {\bibfield
  {journal} {\bibinfo  {journal} {Rev. Mod. Phys.}\ }\textbf {\bibinfo {volume}
  {69}},\ \bibinfo {pages} {629} (\bibinfo {year} {1997})}\BibitemShut
  {NoStop}%
\bibitem [{\citenamefont {Grover}(1978)}]{grover1978}%
  \BibitemOpen
  \bibfield  {author} {\bibinfo {author} {\bibfnamefont {B.~C.}\ \bibnamefont
  {Grover}},\ }\href {\doibase DOI 10.1103/PhysRevLett.40.391} {\bibfield
  {journal} {\bibinfo  {journal} {Phys. Rev. Lett.}\ }\textbf {\bibinfo
  {volume} {40}},\ \bibinfo {pages} {391} (\bibinfo {year} {1978})}\BibitemShut
  {NoStop}%
\bibitem [{\citenamefont {Bear}\ \emph {et~al.}(2000)\citenamefont {Bear},
  \citenamefont {Stoner}, \citenamefont {Walsworth}, \citenamefont
  {Kosteleck\'y},\ and\ \citenamefont {Lane}}]{bear2000}%
  \BibitemOpen
  \bibfield  {author} {\bibinfo {author} {\bibfnamefont {D.}~\bibnamefont
  {Bear}}, \bibinfo {author} {\bibfnamefont {R.~E.}\ \bibnamefont {Stoner}},
  \bibinfo {author} {\bibfnamefont {R.~L.}\ \bibnamefont {Walsworth}}, \bibinfo
  {author} {\bibfnamefont {V.~A.}\ \bibnamefont {Kosteleck\'y}}, \ and\
  \bibinfo {author} {\bibfnamefont {C.~D.}\ \bibnamefont {Lane}},\ }\href
  {\doibase 10.1103/PhysRevLett.85.5038} {\bibfield  {journal} {\bibinfo
  {journal} {Phys. Rev. Lett.}\ }\textbf {\bibinfo {volume} {85}},\ \bibinfo
  {pages} {5038} (\bibinfo {year} {2000})}\BibitemShut {NoStop}%
\bibitem [{\citenamefont {Rosenberry}\ and\ \citenamefont
  {Chupp}(2001)}]{rosenberry2001}%
  \BibitemOpen
  \bibfield  {author} {\bibinfo {author} {\bibfnamefont {M.~A.}\ \bibnamefont
  {Rosenberry}}\ and\ \bibinfo {author} {\bibfnamefont {T.~E.}\ \bibnamefont
  {Chupp}},\ }\href {\doibase 10.1103/PhysRevLett.86.22} {\bibfield  {journal}
  {\bibinfo  {journal} {Phys. Rev. Lett.}\ }\textbf {\bibinfo {volume} {86}},\
  \bibinfo {pages} {22} (\bibinfo {year} {2001})}\BibitemShut {NoStop}%
\bibitem [{\citenamefont {Tullney}\ \emph {et~al.}(2013)\citenamefont
  {Tullney}, \citenamefont {Allmendinger}, \citenamefont {Burghoff},
  \citenamefont {Heil}, \citenamefont {Karpuk}, \citenamefont {Kilian},
  \citenamefont {Knappe-Gr\"uneberg}, \citenamefont {M\"uller}, \citenamefont
  {Schmidt}, \citenamefont {Schnabel}, \citenamefont {Seifert}, \citenamefont
  {Sobolev},\ and\ \citenamefont {Trahms}}]{tullney2013}%
  \BibitemOpen
  \bibfield  {author} {\bibinfo {author} {\bibfnamefont {K.}~\bibnamefont
  {Tullney}}, \bibinfo {author} {\bibfnamefont {F.}~\bibnamefont
  {Allmendinger}}, \bibinfo {author} {\bibfnamefont {M.}~\bibnamefont
  {Burghoff}}, \bibinfo {author} {\bibfnamefont {W.}~\bibnamefont {Heil}},
  \bibinfo {author} {\bibfnamefont {S.}~\bibnamefont {Karpuk}}, \bibinfo
  {author} {\bibfnamefont {W.}~\bibnamefont {Kilian}}, \bibinfo {author}
  {\bibfnamefont {S.}~\bibnamefont {Knappe-Gr\"uneberg}}, \bibinfo {author}
  {\bibfnamefont {W.}~\bibnamefont {M\"uller}}, \bibinfo {author}
  {\bibfnamefont {U.}~\bibnamefont {Schmidt}}, \bibinfo {author} {\bibfnamefont
  {A.}~\bibnamefont {Schnabel}}, \bibinfo {author} {\bibfnamefont
  {F.}~\bibnamefont {Seifert}}, \bibinfo {author} {\bibfnamefont
  {Y.}~\bibnamefont {Sobolev}}, \ and\ \bibinfo {author} {\bibfnamefont
  {L.}~\bibnamefont {Trahms}},\ }\href {\doibase
  10.1103/PhysRevLett.111.100801} {\bibfield  {journal} {\bibinfo  {journal}
  {Phys. Rev. Lett.}\ }\textbf {\bibinfo {volume} {111}},\ \bibinfo {pages}
  {100801} (\bibinfo {year} {2013})}\BibitemShut {NoStop}%
\bibitem [{\citenamefont {Bulatowicz}\ \emph {et~al.}(2013)\citenamefont
  {Bulatowicz}, \citenamefont {Griffith}, \citenamefont {Larsen}, \citenamefont
  {Mirijanian}, \citenamefont {Fu}, \citenamefont {Smith}, \citenamefont
  {Snow}, \citenamefont {Yan},\ and\ \citenamefont {Walker}}]{bulatowicz2013}%
  \BibitemOpen
  \bibfield  {author} {\bibinfo {author} {\bibfnamefont {M.}~\bibnamefont
  {Bulatowicz}}, \bibinfo {author} {\bibfnamefont {R.}~\bibnamefont
  {Griffith}}, \bibinfo {author} {\bibfnamefont {M.}~\bibnamefont {Larsen}},
  \bibinfo {author} {\bibfnamefont {J.}~\bibnamefont {Mirijanian}}, \bibinfo
  {author} {\bibfnamefont {C.~B.}\ \bibnamefont {Fu}}, \bibinfo {author}
  {\bibfnamefont {E.}~\bibnamefont {Smith}}, \bibinfo {author} {\bibfnamefont
  {W.~M.}\ \bibnamefont {Snow}}, \bibinfo {author} {\bibfnamefont
  {H.}~\bibnamefont {Yan}}, \ and\ \bibinfo {author} {\bibfnamefont {T.~G.}\
  \bibnamefont {Walker}},\ }\href {\doibase 10.1103/PhysRevLett.111.102001}
  {\bibfield  {journal} {\bibinfo  {journal} {Phys. Rev. Lett.}\ }\textbf
  {\bibinfo {volume} {111}},\ \bibinfo {pages} {102001} (\bibinfo {year}
  {2013})}\BibitemShut {NoStop}%
\bibitem [{\citenamefont {Korver}\ \emph {et~al.}(2015)\citenamefont {Korver},
  \citenamefont {Thrasher}, \citenamefont {Bulatowicz},\ and\ \citenamefont
  {Walker}}]{kover2015}%
  \BibitemOpen
  \bibfield  {author} {\bibinfo {author} {\bibfnamefont {A.}~\bibnamefont
  {Korver}}, \bibinfo {author} {\bibfnamefont {D.}~\bibnamefont {Thrasher}},
  \bibinfo {author} {\bibfnamefont {M.}~\bibnamefont {Bulatowicz}}, \ and\
  \bibinfo {author} {\bibfnamefont {T.~G.}\ \bibnamefont {Walker}},\ }\href
  {\doibase 10.1103/PhysRevLett.115.253001} {\bibfield  {journal} {\bibinfo
  {journal} {Phys. Rev. Lett.}\ }\textbf {\bibinfo {volume} {115}},\ \bibinfo
  {pages} {253001} (\bibinfo {year} {2015})}\BibitemShut {NoStop}%
\bibitem [{\citenamefont {Limes}\ \emph {et~al.}(2018)\citenamefont {Limes},
  \citenamefont {Sheng},\ and\ \citenamefont {Romalis}}]{limes18}%
  \BibitemOpen
  \bibfield  {author} {\bibinfo {author} {\bibfnamefont {M.~E.}\ \bibnamefont
  {Limes}}, \bibinfo {author} {\bibfnamefont {D.}~\bibnamefont {Sheng}}, \ and\
  \bibinfo {author} {\bibfnamefont {M.~V.}\ \bibnamefont {Romalis}},\ }\href
  {\doibase 10.1103/PhysRevLett.120.033401} {\bibfield  {journal} {\bibinfo
  {journal} {Phys. Rev. Lett.}\ }\textbf {\bibinfo {volume} {120}},\ \bibinfo
  {pages} {033401} (\bibinfo {year} {2018})}\BibitemShut {NoStop}%
\bibitem [{\citenamefont {Donley}\ and\ \citenamefont
  {Kitching}(2013)}]{donley2013}%
  \BibitemOpen
  \bibfield  {author} {\bibinfo {author} {\bibfnamefont {E.}~\bibnamefont
  {Donley}}\ and\ \bibinfo {author} {\bibfnamefont {J.}~\bibnamefont
  {Kitching}},\ }in\ \href@noop {} {\emph {\bibinfo {booktitle} {Optical
  Magnetometry}}}\ (\bibinfo  {publisher} {Cambridge University Press},\
  \bibinfo {year} {2013})\ pp.\ \bibinfo {pages} {369--386}\BibitemShut
  {NoStop}%
\bibitem [{\citenamefont {Walker}\ and\ \citenamefont
  {Larsen}(2016)}]{walker16}%
  \BibitemOpen
  \bibfield  {author} {\bibinfo {author} {\bibfnamefont {T.~G.}\ \bibnamefont
  {Walker}}\ and\ \bibinfo {author} {\bibfnamefont {M.~S.}\ \bibnamefont
  {Larsen}},\ }\href@noop {} {\bibfield  {journal} {\bibinfo  {journal}
  {Advances in Atomic, Molecular, and Optical Physics}\ }\textbf {\bibinfo
  {volume} {65}},\ \bibinfo {pages} {373} (\bibinfo {year} {2016})}\BibitemShut
  {NoStop}%
\bibitem [{\citenamefont {Schaefer}\ \emph {et~al.}(1989)\citenamefont
  {Schaefer}, \citenamefont {Cates}, \citenamefont {Chien}, \citenamefont
  {Gonatas}, \citenamefont {Happer},\ and\ \citenamefont
  {Walker}}]{schaefer1989}%
  \BibitemOpen
  \bibfield  {author} {\bibinfo {author} {\bibfnamefont {S.~R.}\ \bibnamefont
  {Schaefer}}, \bibinfo {author} {\bibfnamefont {G.~D.}\ \bibnamefont {Cates}},
  \bibinfo {author} {\bibfnamefont {T.~R.}\ \bibnamefont {Chien}}, \bibinfo
  {author} {\bibfnamefont {D.}~\bibnamefont {Gonatas}}, \bibinfo {author}
  {\bibfnamefont {W.}~\bibnamefont {Happer}}, \ and\ \bibinfo {author}
  {\bibfnamefont {T.~G.}\ \bibnamefont {Walker}},\ }\href {\doibase DOI
  10.1103/PhysRevA.39.5613} {\bibfield  {journal} {\bibinfo  {journal} {Phys.
  Rev. A}\ }\textbf {\bibinfo {volume} {39}},\ \bibinfo {pages} {5613}
  (\bibinfo {year} {1989})}\BibitemShut {NoStop}%
\bibitem [{\citenamefont {Romalis}\ and\ \citenamefont
  {Cates}(1998)}]{romalis1998}%
  \BibitemOpen
  \bibfield  {author} {\bibinfo {author} {\bibfnamefont {M.~V.}\ \bibnamefont
  {Romalis}}\ and\ \bibinfo {author} {\bibfnamefont {G.~D.}\ \bibnamefont
  {Cates}},\ }\href {\doibase 10.1103/PhysRevA.58.3004} {\bibfield  {journal}
  {\bibinfo  {journal} {Phys. Rev. A}\ }\textbf {\bibinfo {volume} {58}},\
  \bibinfo {pages} {3004} (\bibinfo {year} {1998})}\BibitemShut {NoStop}%
\bibitem [{\citenamefont {Ma}\ \emph {et~al.}(2011)\citenamefont {Ma},
  \citenamefont {Sorte},\ and\ \citenamefont {Saam}}]{ma2011}%
  \BibitemOpen
  \bibfield  {author} {\bibinfo {author} {\bibfnamefont {Z.~L.}\ \bibnamefont
  {Ma}}, \bibinfo {author} {\bibfnamefont {E.~G.}\ \bibnamefont {Sorte}}, \
  and\ \bibinfo {author} {\bibfnamefont {B.}~\bibnamefont {Saam}},\ }\href
  {\doibase 10.1103/PhysRevLett.106.193005} {\bibfield  {journal} {\bibinfo
  {journal} {Phys. Rev. Lett.}\ }\textbf {\bibinfo {volume} {106}},\ \bibinfo
  {pages} {193005} (\bibinfo {year} {2011})}\BibitemShut {NoStop}%
\bibitem [{\citenamefont {Sheng}\ \emph {et~al.}(2014)\citenamefont {Sheng},
  \citenamefont {Kabcenell},\ and\ \citenamefont {Romalis}}]{sheng14a}%
  \BibitemOpen
  \bibfield  {author} {\bibinfo {author} {\bibfnamefont {D.}~\bibnamefont
  {Sheng}}, \bibinfo {author} {\bibfnamefont {A.}~\bibnamefont {Kabcenell}}, \
  and\ \bibinfo {author} {\bibfnamefont {M.~V.}\ \bibnamefont {Romalis}},\
  }\href {\doibase 10.1103/PhysRevLett.113.163002} {\bibfield  {journal}
  {\bibinfo  {journal} {Phys. Rev. Lett.}\ }\textbf {\bibinfo {volume} {113}},\
  \bibinfo {pages} {163002} (\bibinfo {year} {2014})}\BibitemShut {NoStop}%
\bibitem [{\citenamefont {Kwon}\ \emph {et~al.}(1981)\citenamefont {Kwon},
  \citenamefont {Mark},\ and\ \citenamefont {Volk}}]{kwon81}%
  \BibitemOpen
  \bibfield  {author} {\bibinfo {author} {\bibfnamefont {T.~M.}\ \bibnamefont
  {Kwon}}, \bibinfo {author} {\bibfnamefont {J.~G.}\ \bibnamefont {Mark}}, \
  and\ \bibinfo {author} {\bibfnamefont {C.~H.}\ \bibnamefont {Volk}},\ }\href
  {\doibase 10.1103/PhysRevA.24.1894} {\bibfield  {journal} {\bibinfo
  {journal} {Phys. Rev. A}\ }\textbf {\bibinfo {volume} {24}},\ \bibinfo
  {pages} {1894} (\bibinfo {year} {1981})}\BibitemShut {NoStop}%
\bibitem [{\citenamefont {Wu}\ \emph {et~al.}(1987)\citenamefont {Wu},
  \citenamefont {Happer},\ and\ \citenamefont {Daniels}}]{wu1987}%
  \BibitemOpen
  \bibfield  {author} {\bibinfo {author} {\bibfnamefont {Z.}~\bibnamefont
  {Wu}}, \bibinfo {author} {\bibfnamefont {W.}~\bibnamefont {Happer}}, \ and\
  \bibinfo {author} {\bibfnamefont {J.~M.}\ \bibnamefont {Daniels}},\ }\href
  {\doibase DOI 10.1103/PhysRevLett.59.1480} {\bibfield  {journal} {\bibinfo
  {journal} {Phys. Rev. Lett.}\ }\textbf {\bibinfo {volume} {59}},\ \bibinfo
  {pages} {1480} (\bibinfo {year} {1987})}\BibitemShut {NoStop}%
\bibitem [{\citenamefont {Wu}\ \emph {et~al.}(1990)\citenamefont {Wu},
  \citenamefont {Happer}, \citenamefont {Kitano},\ and\ \citenamefont
  {Daniels}}]{wu1990}%
  \BibitemOpen
  \bibfield  {author} {\bibinfo {author} {\bibfnamefont {Z.}~\bibnamefont
  {Wu}}, \bibinfo {author} {\bibfnamefont {W.}~\bibnamefont {Happer}}, \bibinfo
  {author} {\bibfnamefont {M.}~\bibnamefont {Kitano}}, \ and\ \bibinfo {author}
  {\bibfnamefont {J.}~\bibnamefont {Daniels}},\ }\href {\doibase DOI
  10.1103/PhysRevA.42.2774} {\bibfield  {journal} {\bibinfo  {journal} {Phys.
  Rev. A}\ }\textbf {\bibinfo {volume} {42}},\ \bibinfo {pages} {2774}
  (\bibinfo {year} {1990})}\BibitemShut {NoStop}%
\bibitem [{\citenamefont {Slichter}(1990)}]{slichter90}%
  \BibitemOpen
  \bibfield  {author} {\bibinfo {author} {\bibfnamefont {C.~P.}\ \bibnamefont
  {Slichter}},\ }\href@noop {} {\emph {\bibinfo {title} {Principles of Magnetic
  Resonance}}}\ (\bibinfo  {publisher} {Springer},\ \bibinfo {year}
  {1990})\BibitemShut {NoStop}%
\bibitem [{\citenamefont {Sakurai}\ and\ \citenamefont
  {Napolitano}(2013)}]{sakurai2013}%
  \BibitemOpen
  \bibfield  {author} {\bibinfo {author} {\bibfnamefont {J.~J.}\ \bibnamefont
  {Sakurai}}\ and\ \bibinfo {author} {\bibfnamefont {J.}~\bibnamefont
  {Napolitano}},\ }\href@noop {} {\emph {\bibinfo {title} {Modern Quantum
  Mechanics (2nd ed.)}}}\ (\bibinfo  {publisher} {Pearson},\ \bibinfo {year}
  {2013})\BibitemShut {NoStop}%
\bibitem [{\citenamefont {Venema}\ \emph {et~al.}(1992)\citenamefont {Venema},
  \citenamefont {Majumder}, \citenamefont {Lamoreaux}, \citenamefont {Heckel},\
  and\ \citenamefont {Fortson}}]{venema1992}%
  \BibitemOpen
  \bibfield  {author} {\bibinfo {author} {\bibfnamefont {B.~J.}\ \bibnamefont
  {Venema}}, \bibinfo {author} {\bibfnamefont {P.~K.}\ \bibnamefont
  {Majumder}}, \bibinfo {author} {\bibfnamefont {S.~K.}\ \bibnamefont
  {Lamoreaux}}, \bibinfo {author} {\bibfnamefont {B.~R.}\ \bibnamefont
  {Heckel}}, \ and\ \bibinfo {author} {\bibfnamefont {E.~N.}\ \bibnamefont
  {Fortson}},\ }\href {\doibase DOI 10.1103/PhysRevLett.68.135} {\bibfield
  {journal} {\bibinfo  {journal} {Phys. Rev. Lett.}\ }\textbf {\bibinfo
  {volume} {68}},\ \bibinfo {pages} {135} (\bibinfo {year} {1992})}\BibitemShut
  {NoStop}%
\bibitem [{\citenamefont {Donley}\ \emph {et~al.}(2009)\citenamefont {Donley},
  \citenamefont {Long}, \citenamefont {Liebisch}, \citenamefont {Hodby},
  \citenamefont {Fisher},\ and\ \citenamefont {Kitching}}]{donley2009}%
  \BibitemOpen
  \bibfield  {author} {\bibinfo {author} {\bibfnamefont {E.~A.}\ \bibnamefont
  {Donley}}, \bibinfo {author} {\bibfnamefont {J.~L.}\ \bibnamefont {Long}},
  \bibinfo {author} {\bibfnamefont {T.~C.}\ \bibnamefont {Liebisch}}, \bibinfo
  {author} {\bibfnamefont {E.~R.}\ \bibnamefont {Hodby}}, \bibinfo {author}
  {\bibfnamefont {T.~A.}\ \bibnamefont {Fisher}}, \ and\ \bibinfo {author}
  {\bibfnamefont {J.}~\bibnamefont {Kitching}},\ }\href {<Go to
  ISI>://WOS:000262979000124} {\bibfield  {journal} {\bibinfo  {journal} {Phys.
  Rev. A}\ }\textbf {\bibinfo {volume} {79}} (\bibinfo {year}
  {2009})}\BibitemShut {NoStop}%
\bibitem [{\citenamefont {Venema}(1994)}]{venemathesis}%
  \BibitemOpen
  \bibfield  {author} {\bibinfo {author} {\bibfnamefont {B.~J.}\ \bibnamefont
  {Venema}},\ }\href
  {http://www.pqdtcn.com/thesisDetails/7CFF2BB6F3176C4E42781CDFF03A3368}
  {\bibinfo {type} {Doctoral thesis}},\ \bibinfo  {school} {University of
  Washington} (\bibinfo {year} {1994})\BibitemShut {NoStop}%
\bibitem [{\citenamefont {Romalis}\ \emph {et~al.}(1997)\citenamefont
  {Romalis}, \citenamefont {Miron},\ and\ \citenamefont {Cates}}]{romalis1997}%
  \BibitemOpen
  \bibfield  {author} {\bibinfo {author} {\bibfnamefont {M.~V.}\ \bibnamefont
  {Romalis}}, \bibinfo {author} {\bibfnamefont {E.}~\bibnamefont {Miron}}, \
  and\ \bibinfo {author} {\bibfnamefont {G.~D.}\ \bibnamefont {Cates}},\ }\href
  {\doibase 10.1103/PhysRevA.56.4569} {\bibfield  {journal} {\bibinfo
  {journal} {Phys. Rev. A}\ }\textbf {\bibinfo {volume} {56}},\ \bibinfo
  {pages} {4569} (\bibinfo {year} {1997})}\BibitemShut {NoStop}%
\bibitem [{\citenamefont {Volk}\ \emph {et~al.}(1979)\citenamefont {Volk},
  \citenamefont {Mark},\ and\ \citenamefont {Grover}}]{volk1979}%
  \BibitemOpen
  \bibfield  {author} {\bibinfo {author} {\bibfnamefont {C.~H.}\ \bibnamefont
  {Volk}}, \bibinfo {author} {\bibfnamefont {J.~G.}\ \bibnamefont {Mark}}, \
  and\ \bibinfo {author} {\bibfnamefont {B.~C.}\ \bibnamefont {Grover}},\
  }\href {\doibase 10.1103/PhysRevA.20.2381} {\bibfield  {journal} {\bibinfo
  {journal} {Phys. Rev. A}\ }\textbf {\bibinfo {volume} {20}},\ \bibinfo
  {pages} {2381} (\bibinfo {year} {1979})}\BibitemShut {NoStop}%
\bibitem [{\citenamefont {Majumder}\ \emph {et~al.}(1990)\citenamefont
  {Majumder}, \citenamefont {Venema}, \citenamefont {Lamoreaux}, \citenamefont
  {Heckel},\ and\ \citenamefont {Fortson}}]{Majumder1990}%
  \BibitemOpen
  \bibfield  {author} {\bibinfo {author} {\bibfnamefont {P.~K.}\ \bibnamefont
  {Majumder}}, \bibinfo {author} {\bibfnamefont {B.~J.}\ \bibnamefont
  {Venema}}, \bibinfo {author} {\bibfnamefont {S.~K.}\ \bibnamefont
  {Lamoreaux}}, \bibinfo {author} {\bibfnamefont {B.~R.}\ \bibnamefont
  {Heckel}}, \ and\ \bibinfo {author} {\bibfnamefont {E.~N.}\ \bibnamefont
  {Fortson}},\ }\href {\doibase 10.1103/PhysRevLett.65.2931} {\bibfield
  {journal} {\bibinfo  {journal} {Phys. Rev. Lett.}\ }\textbf {\bibinfo
  {volume} {65}},\ \bibinfo {pages} {2931} (\bibinfo {year}
  {1990})}\BibitemShut {NoStop}%
\bibitem [{\citenamefont {Limes}\ \emph {et~al.}(2019)\citenamefont {Limes},
  \citenamefont {Dural}, \citenamefont {Romalis}, \citenamefont {Foley},
  \citenamefont {Kornack}, \citenamefont {Nelson}, \citenamefont {Grisham},\
  and\ \citenamefont {Vaara}}]{limes2019}%
  \BibitemOpen
  \bibfield  {author} {\bibinfo {author} {\bibfnamefont {M.~E.}\ \bibnamefont
  {Limes}}, \bibinfo {author} {\bibfnamefont {N.}~\bibnamefont {Dural}},
  \bibinfo {author} {\bibfnamefont {M.~V.}\ \bibnamefont {Romalis}}, \bibinfo
  {author} {\bibfnamefont {E.~L.}\ \bibnamefont {Foley}}, \bibinfo {author}
  {\bibfnamefont {T.~W.}\ \bibnamefont {Kornack}}, \bibinfo {author}
  {\bibfnamefont {A.}~\bibnamefont {Nelson}}, \bibinfo {author} {\bibfnamefont
  {L.~R.}\ \bibnamefont {Grisham}}, \ and\ \bibinfo {author} {\bibfnamefont
  {J.}~\bibnamefont {Vaara}},\ }\href {\doibase 10.1103/PhysRevA.100.010501}
  {\bibfield  {journal} {\bibinfo  {journal} {Phys. Rev. A}\ }\textbf {\bibinfo
  {volume} {100}},\ \bibinfo {pages} {010501} (\bibinfo {year}
  {2019})}\BibitemShut {NoStop}%
\end{thebibliography}
\end{document}